\documentclass[conference]{IEEEtran}
\IEEEoverridecommandlockouts
\usepackage{xcolor}

\usepackage{textcomp}
\usepackage{xcolor}

\usepackage[normalem]{ulem}
\usepackage[caption=false,font=footnotesize]{subfig}
\usepackage[utf8]{inputenc}
\usepackage{booktabs}
\usepackage{threeparttable}
\usepackage{amsmath,graphicx,cite,amssymb}
\usepackage{amsfonts}
\usepackage{enumitem}
\usepackage{amsthm}
\usepackage{xpatch}
\usepackage{float,cite,amssymb}
\usepackage{graphics}
 \DeclareGraphicsExtensions{.pdf,.jpeg,.pdf,.jpg}   
\usepackage{multirow,bm,bbm,array,setspace}
\usepackage{psfrag}
\usepackage{color}
\usepackage{pstricks}
\usepackage{bbm}
\usepackage{comment}
\usepackage{mdframed}
\usepackage{setspace}
\usepackage{pifont}% http://ctan.org/pkg/pifont
\usepackage{ulem}
\usepackage{booktabs}
\usepackage{array}
\usepackage{arydshln}
\usepackage{algorithm,algpseudocode}

\newcommand{\cS}{\mathcal S}

\newcommand{\bx}{\mathbf{x}}

\newcommand{\bdelta}{\mathbi{\delta}}

\newcommand{\bone}{\mathbf{1}}

\usepackage{bm}
\newcommand{\mathbi}[1]{\bm{#1}}

\newcommand{\cs}{\mathcal{S}}

\newcommand{\ce}{\mathcal{E}}
\newcommand{\cE}{\mathcal{E}}
\newcommand{\cV}{\mathcal{V}}
\newcommand{\cw}{\mathcal{W}}
\newcommand{\cg}{\mathcal{G}}
\newcommand{\bu}{\mathbf{u}}
\newcommand{\bv}{\mathbf{v}}
\newcommand{\R}{\mathbb{R}}

\newcommand{\EE}{\mathbb{E}}

\newtheorem{remark}{Remark}

\renewcommand{\thefigure}{\arabic{figure}}
\begin{document}

\title{Topology Design for Distributed Consensus with Relay-Assisted Communication
\thanks{This work was supported in part by the Knut and Alice Wallenberg Foundation, Swedish Research Council (VR), and Excellence Center at Linköping – Lund in Information Technology (ELLIIT).}
}

\author{\IEEEauthorblockN{Shuyi Ren, Zheng Chen, and Erik G. Larsson}
\IEEEauthorblockA{Dept. of Electrical Engineering (ISY), Linköping University,  Sweden \\
Email: \{shuyi.ren, zheng.chen, erik.g.larsson\}@liu.se
}}

\maketitle

\begin{abstract}

This paper focuses on relay-assisted topology optimization to accelerate the convergence of distributed consensus algorithms over weakly connected networks. Instead of permanently adding a fixed set of relay links, we introduce a time-sharing framework where multiple relay configurations are activated in a probabilistic manner. An ActiveSet-based algorithm incrementally constructs candidate relay edge sets and jointly optimizes the mixing matrices and the associated relay set activation probabilities, allowing adaptive relay selection in the optimization process. Simulation results demonstrate a substantially improved performance–cost trade-off compared with fixed-cardinality relay selection strategies.
\end{abstract}


\section{Introduction}
\label{sec:Intro}

Distributed consensus is a fundamental component of networked systems, allowing nodes to achieve agreement through local information exchange over a communication graph \cite{lynch1996distributed}. The convergence speed of consensus algorithms is strongly influenced by the connectivity of the underlying topology \cite{xiao2004fast,boyd2004fastest}. In many practical system, however, the base topology is constrained by physical limitations such as transmission range, infrastructure development, or network hierarchy. A natural approach to accelerate consensus is therefore to augment the base graph with additional relay-assisted communication links, which effectively improve network connectivity and reduce information propagation delay across the network \cite{growing2006boyd}.

Most existing approaches on topology optimization for distributed consensus select a fixed set of relay edges and permanently augment the base graph~\cite{growing2006boyd,cihan2020topology,summers2015topology}. Although effective, such static designs may not fully exploit relay-assisted communication under cost constraints. Instead, one can adopt a time-sharing strategy in which multiple relay configurations are activated over time, either according to a deterministic schedule, or randomly according to a probability distribution \cite{hovareshti2007probabilistic,boyd2006randomized,zhang2025timevaryingmixingmatrixdesign,griparic2022consensus}. In this work we focus on probabilistic activation, where configurations are selected randomly in each iteration~\cite{daniel2025faster,wang2019matcha, 10038615}. This mechanism enlarges the design space and can improve the consensus performance even when individual configurations are relatively simple.

Furthermore, the benefit of adding relay links typically exhibits diminishing returns, while the communication cost increases with the number of activated edges. Fixing the number of relay links therefore limits the achievable performance–cost trade-off. To overcome this limitation, we propose a topology design method, termed \textit{ActiveSet}, which adaptively constructs relay configurations and determines their activation probabilities.

The main contributions of this work are threefold. First, we introduce a time-sharing topology design framework enabling probabilistic activation of multiple relay configurations. Second, we propose the ActiveSet algorithm that incrementally constructs relay edge sets and jointly optimizes their mixing matrices and activation probabilities. Third, simulations on clustered graphs with bridge-and-tail structures demonstrate significant improvements in consensus convergence compared with fixed-cardinality relay augmentation methods.

\section{System Model and Preliminaries}
\label{sec:model}

We consider a network with the base communication topology modeled as a graph $(\cV,\cE_0)$, where $\cV=\{1,\ldots,N\}$ denotes the set of nodes and $\cE_0$ denotes the set of links with $|\cE_0|=E$.  The graph models the default communication structure determined by physical constraints such as transmission range or infrastructure deployment. The connectivity between nodes is characterized by the adjacency matrix $A=[a_{ij}]$, where $a_{ij}=1$ if $(i,j)\in\mathcal{E}_0$ and $a_{ij}=0$ otherwise. The graph Laplacian is defined as
\begin{equation}
    L =D - A,
\end{equation}
where $D=\mathrm{diag}(d_1,\ldots,d_N)$ is the degree matrix with $d_i=\sum_{j=1}^N a_{ij}$.

The goal of the system is to achieve average consensus, i.e., all nodes agree asymptotically on the average of their initial values by following an iterative averaging protocol. Let $x_i(t)$ denote the state held by node $i$ at iteration $t$. The standard linear consensus update is  
\begin{equation}
    \bx(t+1)=W\bx(t),
\end{equation}
where $\bx(t)=[x_1(t),\ldots,x_N(t)]^\top$ and $W\in\mathbb{R}^{N\times N}$ is the mixing matrix, which we assume satisfies: (i) non-negativity and sparsity consistent with the communication graph, i.e., $W_{ij}>0$ only if $(i,j)\in\cE_0$ or $i=j$; (ii) $W$ can be asymmetric but we assume it is doubly stochastic, i.e., $W\bone = \bone$ and $\bone^\top W = \bone^\top$.

Under these conditions, the network average is preserved, i.e., $\frac{1}{N}\bone^\top \bx(t)=\frac{1}{N}\bone^\top \bx(0),\;\forall t$. And, under standard primitivity conditions, the linear
iteration converges to the initial average $\bx^\star=J\bx(0)$, where $J:=\frac{1}{N}\bone\bone^T$. 

Furthermore, the disagreement  $\bdelta_t:=\bx(t)-J\bx(t)$ evolves as
\begin{equation}
  \bdelta_{t+1}=(W-J)\bdelta_t.
\end{equation}
%Thus, the convergence speed is determined by the contraction factor $\|W-J\|_2$. 
Thus, the asymptotic rate is governed by $\rho(W-J)$~\cite{xiao2004fastLinear}, while $\|W-J\|_2$ provides a tractable upper bound. The two quantities coincide when $W$ is symmetric, but may differ for asymmetric $W$.

Note that the second smallest eigenvalue $\lambda_2(L)$, known as the algebraic connectivity, is a classical measure of graph connectivity. It is well known that $\lambda_2(L)$ is monotone non-decreasing with respect to edge additions, i.e., adding edges cannot decrease $\lambda_2(L)$. Relay edges can accelerate consensus by improving the contraction of the mixing matrix on the disagreement subspace. In our setting, the asymptotic consensus rate is characterized by $\rho(W-J)$, while the algorithm minimizes the tractable upper bound $\|W-J\|_2$. This observation motivates the use of relay-assisted links to strengthen bottleneck structures such as bridges and tail nodes, which we formalize in the following model.

\begin{remark}[Asymmetric mixing matrices]\label{rem:asym}
In relay-assisted systems, communication and forwarding can be directional and heterogeneous; allowing asymmetric $W$ capturing directed information flow and scheduling effects.
\end{remark}

\vspace{-0.2cm}
\subsection{Relay-Assisted Time-Sharing Topology Augmentation}

A candidate relay edge represents a potential link that can be activated to enhance the communication topology. In wireless systems, this may correspond to repeaters, RIS-assisted links, or backhaul connections. 

At each iteration, $K$ relay edges are selected from a set of possible relay edges $\cE_{\rm cand}$ with cardinality $|\cE_{\rm cand}|=R$. Define $\cS_K=\{\cE^{(1)},\ldots,\cE^{(M)}\}$ as the collection of all candidate relay sets, where $\cE^{(m)}\subseteq \cE_{\rm cand}$, $|\cE^{(m)}|=K$ and $M=\binom{R}{K}$. For any selected relay set $\cE^{(m)}$, the augmented topology becomes $\cE_0\cup \cE^{(m)}$. Let $\gamma_{\cE}$ denote the communication cost associated with activating a relay set $\cE$, and let $\Gamma$ represent the average communication budget. 
A detailed communication model will be introduced later in the simulation section. At each iteration $t$, one configuration index $m_t\in\{1,\ldots,M\}$ is sampled according to a probability distribution $P=(p_1,\ldots,p_M)$ with $p_m\ge 0$ and $\sum_{m=1}^M p_m=1$. The consensus update is then  $$\bx_{t+1}=W_{m_t}\bx_t,$$
where $W_{m_t}\in\cw(\cE^{(m_t)})$. For any relay set $\cE$, we define 
\begin{equation}
    \begin{split}
        \cw(\cE):=\Big\{W\in\R^{N\times N}:\ 
W\bone=\bone,\ 
\bone^\top W=\bone^\top,\ \\
W\ge0,\ 
[W]_{ij}=0\ \text{if}\ (i,j)\notin \cE_0\cup\cE
\Big\}.
    \end{split}
\end{equation}

\subsection{Static vs. Dynamic Relay Activation Strategies}
\label{subsec:baseline}

Before introducing the proposed topology augmentation method, we present some comparative examples to illustrate the limitations of static relay activation strategies, which motivates the proposed time-sharing dynamic activation design. For performance comparison, We focus on the type of networks that contain two weakly connected clusters, connected by a limited number of bridge links and possibly several tail nodes. 

\subsubsection{Adding fixed set of edges (Static-TopK)}
As proposed in \cite{growing2006boyd}, a fundamental approach is to select a fixed set of $K$ edges that maximizes the algebraic connectivity of the augmented graph. Formally, this corresponds to solving
$
\max_{\cE\subseteq \cE_{\rm cand},\,|\cE|=K} \ \lambda_2\big(L_0 + L(\cE)\big),
$
where $L_0$ denotes the Laplacian of the baseline graph and $L(\cE)$ is the Laplacian contribution of the added edges. The resulting relay set $\cE$ is then permanently activated at every iteration. The main limitations of this design are: 
1) it enforces a fixed number of relays; 2) it uses the same relay set at every iteration, limiting the ability to exploit diverse relay configurations.

\subsubsection{Gained-guided Sampling (GainGuided)} This design
is inspired by the static method in \cite{growing2006boyd}. For each candidate edge $r\in\{1,\ldots,R\}$, define its gain as the increase in algebraic connectivity when the edge is added individually, $\mathrm{gain}_r=\lambda_2(L_0+L_r)-\lambda_2(L_0)$, where $L_0$ denotes the Laplacian of the base graph and $L_r$ represents the Laplacian contribution of edge $r$. An edge level probability can be constructed as $\hat{p}_r=\mathrm{gain}_r/\sum_{j=1}^R\mathrm{gain}_j$, $\forall r=1,\ldots,R$. A relay set in $\cS_K$ is then sampled according to the induced distribution $P=(p_1,\ldots,p_M)$ where $p_m=\sum_{\pi\in\text{Perm}(\cE^{(m)})}\prod_{k=1}^K\frac{\hat{p}_{\pi_k}}{1-\sum_{j=1}^{k-1}\hat{p}_{\pi_j}}$. While this heuristic favors edges with strong individual gains, it ignores joint interactions among edges.

\subsubsection{Optimal fixed-$K$ design (JointOpt-K)} A more principled approach jointly optimizes the activation distribution $P=(p_1,\ldots,p_M)\in\Delta_M$ and the mixing matrices $\{W_m\}_{m=1}^M$. Conditioning on $\bx_t$ yields 
\[ 
\EE[\bx_{t+1}\mid \bx_t]=\bar W(P)\bx_t,\qquad \bar W(P):=\sum_{m=1}^M p_m W_m. 
\] 
Since each $W_m$ is doubly stochastic, $\bar W(P)$ is also doubly stochastic.
By the Birkhoff-von Neumann theorem~\cite{birkhoff1946tres}, $\bar W(P)$ is a convex combination of
permutation matrices. Permutation matrices are orthogonal, and therefore they have spectral norm
equal to one. By convexity of the spectral norm~\cite{horn2012matrix}, it follows that $\|\bar W(P)\|_2 \le 1$. Together with $\bar W(P)-J=(I-J)\bar W(P)(I-J)$, this yields $\|\bar W(P)-J\|_2\le 1$.  

Since $\EE[\bx_{t+1}\mid \bx_t]=\bar W(P)\bx_t$ and $\bar W(P)J=J$, the conditional mean disagreement evolves as
$$\EE[\bdelta_{t+1}] = (\bar W(P)-J)\EE[\bdelta_{t}],$$ where $\bdelta_t= \bx(t) - J\bx(t)$. Therefore,
$\|\EE[\bdelta_{t+1}]\|_2
\le \|\bar W(P)-J\|_2\,\|\EE[\bdelta_{t}]\|_2.$
Thus, $\|\bar W(P)-J\|_2$ characterizes the one-step contraction of the
conditional mean disagreement. Since $\rho(\bar W(P)-J)\le \|\bar W(P)-J\|_2$, minimizing the spectral norm provides a tractable surrogate for accelerating consensus under randomized topology switching. Moreover, if the attained value of $\|\bar W(P)-J\|_2$ is strictly less than $1$, it also certifies geometric convergence of the expected disagreement. On the other hand, if the attained value equals $1$, the objective is still meaningful as a surrogate, even though the spectral norm does not characterize the actual convergence rate.

For a fixed integer $K$, we consider the following problem
\begin{equation} \label{opt:fixed_library_obj} 
\min_{\{W_m\},\,P}\quad \big\|\bar W(P)-J\big\|_2 
\end{equation} 
subject to the communication budget constraint $\mathbb{E}[\gamma_{\cE}]\leq\Gamma$, where $\cE^{(m)}\in \cS_K$, $W_m\in \cw(\cE^{(m)})$, and $P\in\Delta_M$. 
The problem can be solved by using alternating optimization \cite{daniel2025faster}.

\begin{figure}[t!]
    \vspace{-1em}
    \centering
    \includegraphics[width=\linewidth]{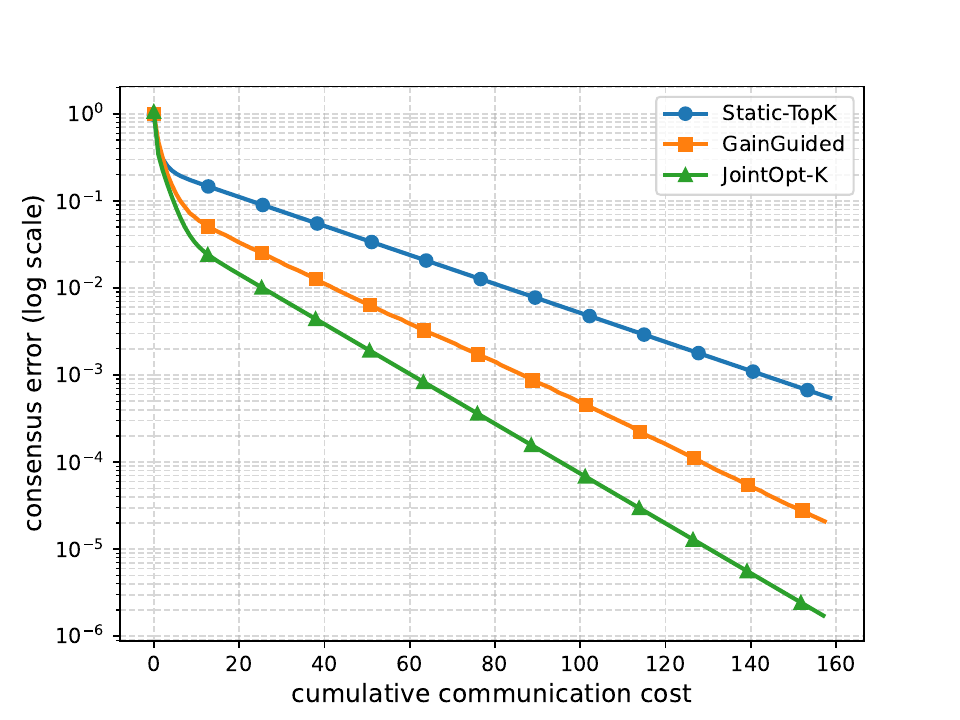}
    \caption{Illustration of the effects of adding fixed size relay set}
    \label{fig:preliminary}
    \vspace{-2.0em}
\end{figure}

Fig.~\ref{fig:preliminary} illustrates the evolution of the consensus error $\|\bx_t-\bar{x}\bone\|_2$ versus the number of iterations for the aforementioned relay selection strategies. We observe that random time-sharing among relay configurations can significantly improve convergence rate compared to the other schemes. However, it suffers from a fundamental limitation. The number of feasible relay configurations grows combinatorially with $K$, since $M=\binom{R}{K}$, making the design space intractable as $K$ increases. Moreover, fixing $K$ restricts the achievable performance-cost trade-off and enforces a uniform relay budget across activations, which may be suboptimal for heterogeneous network topologies.

Motivated by these limitations, we develop an active set relay topology design framework that jointly determines relay set configurations, their cardinalities, the corresponding mixing matrices $\{W_m\}$, and the activation probabilities $P$, with the goal of optimizing consensus performance under communication cost constraints.
%\vspace{-0.2em}

\section{ActiveSet Relay Topology Design}
\label{sec:algorithm}

We propose an active set based topology design framework that constructs a library of relay configurations adaptively. Instead of enumerating the entire configuration library $\cS_K$, the proposed method incrementally constructs a small active library of promising relay configurations. At each iteration, a restricted joint optimization problem is solved over the current library to obtain mixing matrices and activation probabilities. The resulting solution is then used to guide the expansion of the library via a dual-informed edge pricing mechanism. In this way, relay configurations and their cardinalities emerge adaptively from the optimization process, avoiding the combinatorial explosion associated with fixed-$K$ enumeration.

\subsection{Dual-Guided Edge Pricing}
Recall from Section~\ref{sec:model} that the expected mixing matrix under the time-sharing model is $\bar W(P)=\sum_{m=1}^M p_m W_m$, and the consensus contraction is
\begin{equation}\label{eq:disagreement}
    f(B)=\|B\|_2, \qquad B:=\bar W(P)-J .
\end{equation}
Let the singular value decomposition of $B$ be $
B=U\Sigma V^\top$, and denote by $\bu$ and $\bv$ the dominant left and right singular vectors, respectively.  When the largest singular value is simple, a subgradient of $f(B)=\|B\|_2$ is given by \cite{watson1992characterization,lewis1995convex}
\begin{equation}\label{eq:est_grad}
G=\bu\bv^\top \in \partial \|B\|_2 .
\end{equation}
If the largest singular value is not simple, $\|B\|_2$ is generally
nondifferentiable, and the subdifferential is given by the convex hull of outer products formed by singular vector pairs associated with $\sigma_{\max}(B)$. For simplicity, we use a dominant singular vector pair returned by the SVD routine as a pricing direction. For a perturbation $\Delta B$,
\[
\|B+\Delta B\|_2\approx\sigma_{\max}(B)+\langle G,\Delta B\rangle,
\]
where $\langle G,\Delta B\rangle=\mathrm{trace}(G^\top \Delta B)=\sum_{i,j}G_{ij}[\Delta B]_{ij}$. This relation provides a first-order sensitivity of the contraction factor with respect to perturbations of $\bar{W}$, which we use to guide the search for beneficial relay edges.

To preserve double stochasticity under asymmetric mixing matrices, we consider a perturbation direction $\Delta W$ satisfying $\Delta W\bone=0$ and $ \bone^\top\Delta W=0$. Specifically,
\[
\Delta W^{(i,j;r,c)}=\varepsilon(E_{ij}-E_{ic}-E_{rj}+E_{rc}),
\]
where $E_{ab}$ denotes the matrix with a $1$ at position $(a,b)$ and $0$ elsewhere, and $0<\varepsilon\ll1$ is a small step size. This construction preserves the row and column sums, and preserves non-negativity provided $\varepsilon$ is sufficiently small and the decremented entries satisfy $W_{ic}\ge\varepsilon$ and $W_{rj}\ge\varepsilon$.

For a candidate relay edge $(i,j)$, we define the predicted improvement
\[
s(i,j|\ce)=\min_{(r,c)\in\mathcal{F}(i,j|\ce)}(\bu[i]-\bu[r])(\bv[j]-\bv[c]),
\]
where $\mathcal{F}(i,j|\ce)$ denotes the set of feasible swap partners that preserve the sparsity pattern of $\cE_0\cup\ce$ and maintain non-negativity of the perturbed mixing matrix for sufficiently small $\varepsilon$. More negative values indicate stronger potential reduction of the contraction factor. Incorporating the communication cost $\gamma_{(i,j)}$, the net gain of adding edge $(i,j)$ to parent configuration $\cE$ is given as
\begin{equation}\label{eq:gain_net}
\cg(i,j|\cE):= -s(i,j)-\gamma_{(i,j)}.
\end{equation}

\subsection{ActiveSet Library Expansion}
The algorithm maintains an active configuration library $\cs_{\rm active}$. The library $\mathcal{S}_{\rm active}$ is initialized with all single-edge relay configurations. At each iteration, we first solve the restricted optimization problem \eqref{opt:fixed_library_obj} over the current library to obtain the mixing matrices $\{W(\ce)\}$ and activation probabilities $P$. The resulting solution is then used to guide the expansion and pruning of the library. The library is initialized with single-edge relay sets. The
following steps are performed in each iteration.

\textit{Parent Freezing}: For each configuration $\cE\in\cS_{\text{active}}$, compute
\[
\cg_{\max}(\cE):=\max_{(i,j)\notin\cE}\;\cg(i,j|\cE).
\]
If $\cg_{\max}(\cE)\leq 0$, the configuration is not expanded.

\textit{Expansion}: For expandable parents, compute gains for candidate edges and generate a shortlist of children 
\[
\cE'=\cE\cup\{(i,j)\}.
\]
Only the top $Q$ edges are retained, where $Q$ is a user-defined parameter that controls the number of candidate expansions per parent and balances exploration and computational complexity.

\textit{Duplicate removal and pruning}: Children generated from different parents may coincide. Configurations are stored using a canonical representation, and duplicate entries are merged. After solving the restricted optimization problem \eqref{opt:fixed_library_obj}, configurations with small activation probability $p(\cE)<\tau$, where $\tau$ is a prescribed threshold, are removed and the remaining probabilities are renormalized. This pruning step keeps the active library compact and prevents the algorithm from expanding configurations deemed irrelevant by the optimizer.

The proposed method follows an ActiveSet strategy reminiscent of
column-generation methods [14]: a restricted master problem optimizes the current configuration library, while a pricing step identifies relay configurations that can further reduce the objective. The algorithm iteratively expands the relay configuration library using a dual-guided pricing rule while maintaining a compact active set through pruning.
The complete procedure is summarized in Algorithm \ref{alg:relay_active_set}.

\begin{algorithm}[h!]
\caption{ActiveSet Relay Topology Design}
\label{alg:relay_active_set}
\begin{algorithmic}
\Require $(\cV,\cE_0)$, $\mathcal{E}_{\rm cand}$, $\{\gamma_e\}$
\State \textbf{Initialize:} active library $\mathcal{S}_{\rm active}$ with all single-edge sets
\State Set thresholds $\kappa$, $\tau$ and shortlist size $Q$
\Repeat
    \State \textbf{Restricted master problem}
    \State Solve \eqref{opt:fixed_library_obj} over $\mathcal{E}\in\mathcal{S}_{\rm active}$ to obtain $\{W(\mathcal{E})\}$ and $P$
    
    \State Prune configurations with $p(\mathcal{E})<\tau$ and renormalize $P$
    
    \State Update $\bar{W}(P)$ and compute $B\gets \bar{W}(P)-J$ by~\eqref{eq:disagreement}
    \State Compute dominant singular vectors, $G\gets \bu\bv^\top$ by~\eqref{eq:est_grad}
    
    \State \textbf{Pricing step}
    \State Initialize candidate pool $\mathcal{C}\gets\emptyset$
    
    \ForAll{$\mathcal{E}\in\mathcal{S}_{\rm active}$}
        \ForAll{$(i,j)\in\mathcal{E}_{\rm cand}\setminus\mathcal{E}$}
            \State Compute gain $\mathcal{G}(i,j\mid\mathcal{E})$ using \eqref{eq:gain_net}
            \If{$\mathcal{G}(i,j\mid\mathcal{E})>\kappa$}
                \State Add child $\mathcal{E}\cup\{(i,j)\}$ to $\mathcal{C}$
            \EndIf
        \EndFor
    \EndFor
    
    \State Remove duplicate relay sets in $\mathcal{C}$
    \State Add top-$Q$ children from $\mathcal{C}$ into $\mathcal{S}_{\rm active}$

\Until{no child satisfies $\mathcal{G}>\kappa$}
\State \Return $\mathcal{S}_{\rm active}$, $\{W(\mathcal{E})\}$, $P$
\end{algorithmic}
\end{algorithm}

\subsection{Computational Complexity}
Let $|\mathcal{S}_{\rm active}|$ denote the size of the active configuration library. Each iteration first solves the restricted problem \eqref{opt:fixed_library_obj} over the current library, whose cost scales with $|\mathcal{S}_{\rm active}|$. The dominant singular vectors of $B=\bar W(P)-J$ are then computed, which can be obtained efficiently using power iterations with complexity $\mathcal{O}(N^2)$ per iteration.

The pricing step evaluates candidate edges $(i,j)\in\mathcal{E}_{\rm cand}\setminus\cE$ for each parent $\cE\in\mathcal{S}_{\rm active}$. Computing the full score $s(i,j\mid\cE)$ in \eqref{eq:gain_net} requires searching over feasible swap partners $(r,c)$, which can be costly when the candidate set is large. To reduce this overhead, we may use the surrogate score $\tilde{s}(i,j\mid\cE)=|\bu[i]-\bu[j]|\cdot|\bv[i]-\bv[j]|$, which avoids enumerating all swap pairs and significantly reduces the pricing complexity in practice.
Finally, pruning configurations with small activation probability
keeps $|\mathcal{S}_{\rm active}|$ small and prevents the library from growing combinatorially.

\section{Simulation}
\label{sec:simulation}

To model communication costs, we assign each relay edge $(i,j)$ a distance-based cost
\[
\gamma_{ij}=\gamma_0+\alpha\log(1+(d_{ij}/d_0)^k)
\]
where $d_{ij}$ denotes the Euclidean distance between nodes $i$ and $j$, $d_0$ is a normalization length, and $k$ plays a role similar to a path-loss exponent. This model reflects the fact that longer communication links typically require higher transmission power or reduced communication rates.

We evaluate the proposed method on two types of baseline graphs. In both cases, the ActiveSet method—which adaptively constructs relay configurations without fixing $K$—is compared with the baseline methods introduced in Section~\ref{subsec:baseline}. Furthermore, to highlight the importance of optimizing the activation distribution $P$, we consider an additional baseline
 \textit{Uniform}, which performs uniform sampling over all relay configurations.

\paragraph{Heterogeneous Relay Opportunity Graph} We first consider a topology with heterogeneous relay opportunities. In this setting, one candidate relay edge provides a high-impact but expensive shortcut between clusters, while several cheaper relay edges offer local connectivity improvements. Such heterogeneity makes fixed-$K$ augmentation inefficient, since all activations must use the same number of relay edges.
 
\begin{figure}[t!]
\centering
\includegraphics[width=\linewidth]{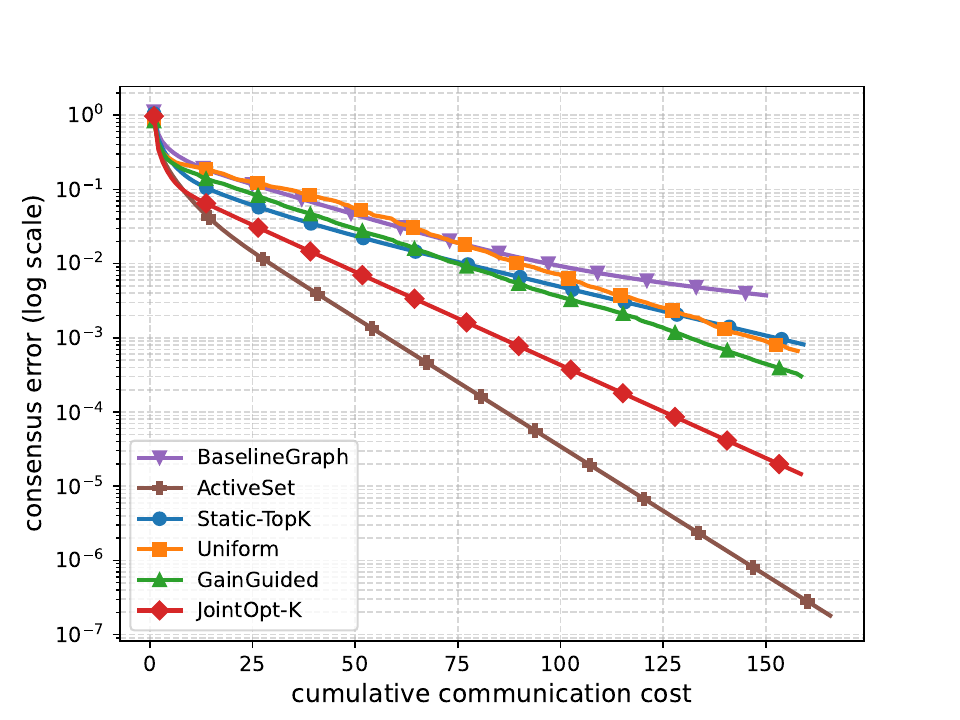}
\includegraphics[width=\linewidth]{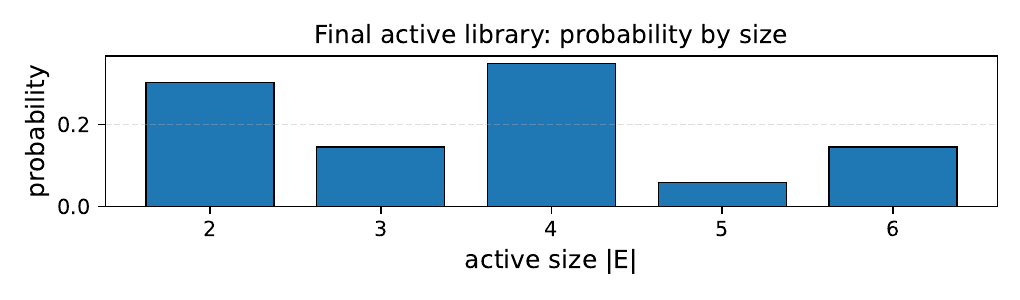}
\caption{Results of heterogeneous relay opportunity graph.}
\label{fig:hetero_result}
\vspace{-1.0em}
\end{figure}

The top panel in Fig.~\ref{fig:hetero_result} shows the consensus error as a function of communication cost. The proposed ActiveSet method outperforms all fixed-$K$ strategies. This improvement arises because the algorithm learns to occasionally activate the expensive shortcut while more frequently selecting cheaper local relay edges. In contrast, fixed-$K$ designs cannot adapt the relay cardinality to the vary benefit of different relay opportunities. 
To illustrate the behavior of the proposed method, the bottom panel in Fig.~\ref{fig:hetero_result} shows the distribution of relay set sizes selected by the ActiveSet algorithm. The learned policy mixes configurations with different cardinalities, occasionally activating larger relay sets when the expected improvement justifies the additional cost. This adaptive behavior explains the performance advantage over fixed-$K$ designs.

\paragraph{Random Bridged-Cluster Graph} We next evaluate the algorithms on randomly generated graphs consisting of two Erd\H{o}s–R\'{e}nyi clusters connected by a single bridge edge. To create additional local bottlenecks, several tail paths are attached to randomly selected nodes within each cluster.

\begin{figure}[t!]
    \centering
    \includegraphics[width=\linewidth]{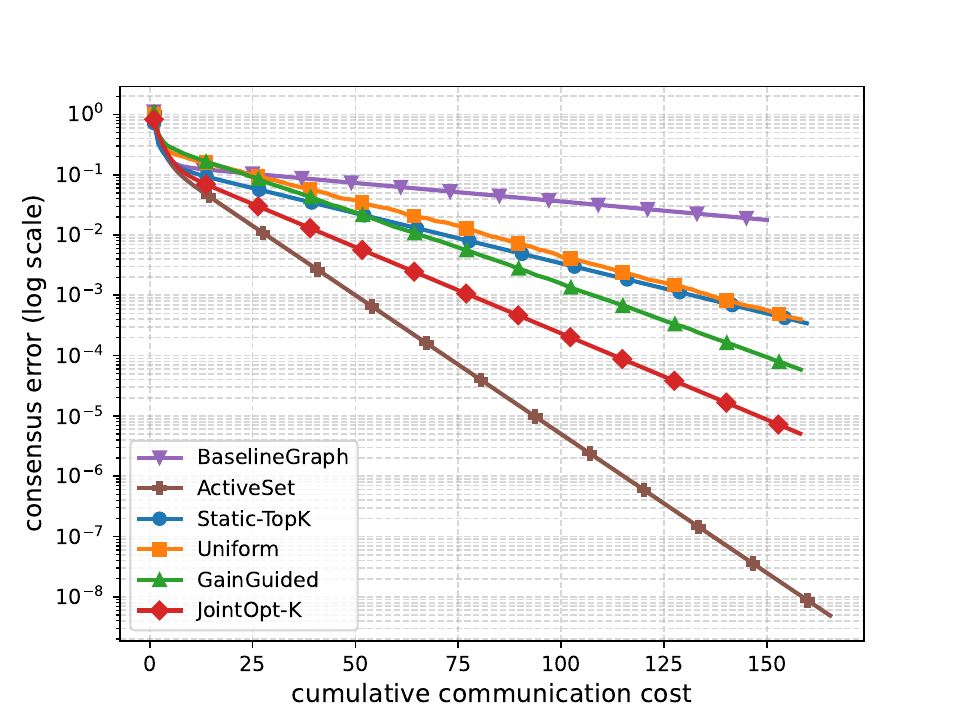}
   % \vspace{-2.5em}
    \caption{Results of randomly generated bridged-cluster graph}
   % \vspace{-1em}
    \label{fig:active_set}
    \vspace{-2.0em}
\end{figure}

Fig.~\ref{fig:active_set} shows the resulting consensus performance. Similar trends are observed: the proposed ActiveSet approach consistently achieves faster convergence for a given communication cost. The improvement is particularly pronounced in the presence of tail structures, where relay edges near the bottleneck bridge and tail roots provide the largest performance gains.

\section{Conclusion}
\label{sec:con}
This paper studied relay-assisted topology design for accelerating
distributed consensus in weakly connected networks. A time-sharing framework with probabilistic activation
of relay configurations was introduced, together with an ActiveSet
algorithm that jointly optimizes the relay activation
probabilities and their associated mixing matrices  under communication budget constraints. Simulation results
show that the proposed method significantly outperforms fixed-cardinality
topology augmentation strategies, highlighting the effectiveness of
adaptive relay configuration design for improving consensus convergence.

\bibliographystyle{IEEEtran}     
\bibliography{strings, references}

\end{document}